\documentclass[
  aps,
  prb,
  reprint,
  superscriptaddress,
  longbibliography,
  floatfix
]{revtex4-2}

\usepackage{graphicx}
\usepackage{amssymb}
\usepackage{dcolumn}
\usepackage{color}
\usepackage{bm}
\usepackage{amsmath}
\usepackage{braket}
\usepackage{siunitx}
\usepackage[version=4]{mhchem}
\usepackage{hyperref}
\hypersetup{
  colorlinks=true,
  linkcolor=blue,
  citecolor=blue,
  urlcolor=blue,
  pdftitle={PRB Skeleton Title},
  pdfauthor={First A. Author}
}
\usepackage{cleveref}
\crefname{appendix}{App.}{Apps.}
\crefname{equation}{Eq.}{Eqs.}
\crefname{figure}{Fig.}{Figs.}
\crefname{table}{Tab.}{Tabs.}
\crefname{section}{Sec.}{Secs.}

\graphicspath{{figures/}}

\begin{document}

\title{Bootstrap bounds for Quantum Spin Systems using String Operators}
\author{Nisarg Chadha}
\affiliation{Department of Physics, Harvard University, Cambridge, Massachusetts 02138, USA}
\author{Michael G. Scheer}
\affiliation{Department of Physics, Harvard University, Cambridge, Massachusetts 02138, USA}
\author{Eslam Khalaf}
\affiliation{Department of Physics, Harvard University, Cambridge, Massachusetts 02138, USA}
\date{\today}
\begin{abstract}

Bootstrap is a numerical many-body method that provides rigorous bounds on ground-state observables by imposing a set of necessary constraints on the expectation values of operators. The quality of the resulting bounds is sensitive to the choice of operators entering the constraints. In particular, bounds on ground-state correlations are often loose in spontaneous symmetry-breaking (SSB) phases, since local operator sets cannot exclude domain-wall excitations. In this work, we introduce non-local, string-like operators into the bootstrap and show that the program can be formulated directly in thermodynamic limit. We then apply our construction to several 1D spin models. First, we obtain a significant tightening of the bounds in the SSB phase of the 1D transverse-field Ising model. Using the 1D axial next-nearest-neighbor Ising model, we further show that this tightening allows for a quantitative estimate of the locations of phase boundaries. Finally, we generalize the string operators to the 1D $\mathbb{Z}_3$ chiral clock model and track the behavior of the bounds across the phase diagram. Our results broaden the class of constraints available to the bootstrap and open a route toward bootstrapping more general symmetry-broken and topological phases, where the relevant constraints may involve non-local or extended operators.
\end{abstract}

\maketitle
\section{Introduction}
Computing ground-state observables for interacting systems is a key problem in many-body physics. The exponential growth of the Hilbert space dimension with system size renders exact numerical solutions infeasible for large systems. This difficulty is compounded by the fact that we are mostly interested in the thermodynamic limit, where the system size is taken to infinity.

To address this, variational methods such as density functional theory \cite{RevModPhys.61.689,burke2012perspective}, Hartree-Fock theory \cite{echenique2007mathematical}, tensor network methods \cite{verstraete2008matrix, PhysRevLett.101.110501,PhysRevB.80.235127, RevModPhys.93.045003}, and neural quantum states \cite{carleo2017solving,lange2024architectures} have been used successfully to estimate ground-state observables for a wide range of interacting systems.
However, methods based on the variational principle offer no certificate or guarantee on the accuracy of correlators, beyond the fact that the variational energy provides an upper bound on the true ground-state energy.

Recently, an approach known as the \emph{Hamiltonian bootstrap} or \emph{many-body bootstrap} has been introduced as a method to obtain rigorous two-sided bounds on thermodynamic ground-state (and finite-temperature Gibbs state) observables \cite{fawzi2024certified, Cho2026CoarseGrainedBootstrap, cho2025thermal, scheer2024hamiltonian}. The approach relies on the idea of relaxation: one imposes a set of necessary but not sufficient constraints on ground-state correlations. Rigorous bounds are then obtained by optimizing over this enlarged feasible set.

Relaxation methods were first introduced by Anderson  \cite{PhysRev.83.1260}, who derived a lower bound on the ground-state energy density of the 1D Heisenberg antiferromagnet. By expressing the Hamiltonian as a sum of simpler sub-Hamiltonians, Anderson lower bounded the full ground-state energy by the sum of the ground-state energies of the sub-Hamiltonians.
This method was subsequently applied to other frustrated spin systems to obtain reasonably good lower bounds \cite{majumdar1976lower,PhysRevB.41.9611}.

A related line of development in the quantum chemistry community began with Coleman's observation that the ground-state problem for a Hamiltonian with two-body interactions could be written as an optimization of a linear functional of the two-body reduced density matrix (2-RDM) over the space of all 2-RDMs compatible with an $N$-particle quantum state \cite{RevModPhys.35.668}.
The set of compatibility conditions parametrizing this space came to be known as the $N$-representability constraints, and this idea was further developed in Ref. \cite{garrod1964reduction}.
Later work developed tractable necessary conditions for $N$-representability and formulated them as semidefinite programs (SDPs), yielding a hierarchy of relaxations whose optima provide rigorous lower bounds on the ground-state energy \cite{nakata2001variational,PhysRevA.65.062511, PhysRevLett.93.213001, PhysRevLett.108.263002}.
The relaxation of the $N$-representability constraints has also been applied to lower bound ground-state correlations in strongly interacting condensed matter systems \cite{csnn-vjhn, gw85-5r92}.

More recently, the ground-state relaxation problem has been recast in terms of positive semidefinite constraints on moment matrices \cite{PhysRevLett.108.200404,baumgratz2012lower,han2020quantum,fawzi2024certified, Cho2026CoarseGrainedBootstrap,scheer2024hamiltonian}, bringing it closer to the operator-algebraic formulation of quantum systems \cite{Bratteli1987, Bratteli1997}.
Identifying additional constraints specific to the ground-state density matrix has enabled the development of bootstrap methods that provide rigorous two-sided bounds on any local ground-state correlation, even in the thermodynamic limit \cite{Cho2026CoarseGrainedBootstrap,scheer2025defect}.

In practice, however, these bounds can be quite loose, as observed in the ferromagnetic (FM) phase of the 1D transverse-field Ising model (TFIM) \cite{fawzi2024certified}.
Ref. \cite{Cho2026CoarseGrainedBootstrap} implemented a coarse-graining scheme introduced in Ref. \cite{PhysRevX.14.031006} to incorporate a considerably larger number of constraints.
However, the bounds in the FM phase of the 1D TFIM only improved marginally despite coarse-graining to relatively larger operator sets.

In a recent paper \cite{scheer2025defect}, we identified the key obstacle to bootstrapping spontaneous symmetry-broken (SSB) phases to be the presence of proliferating domain-wall excitations.
A finite bootstrap program built from local operators is unable to rule out states with large domain walls from the set of feasible solutions, which leads to loose bounds.
In Ref. \cite{scheer2025defect}, we addressed this by enlarging the Hilbert space with auxiliary defect variables, allowing us to locally exclude domain walls and thereby significantly improve the bounds on ground-state correlations.

This Hilbert space enlargement procedure does, however, have a few drawbacks. First, one must show that ground-state bounds in the enlarged model carry over to ground-state bounds in the original problem. This can be non-trivial and may require special properties of the model. Additionally, the Hilbert space enlargement complicates the generalization of the method to finite temperature.

In this paper, we introduce an alternative route for incorporating domain-wall excitations that directly employs non-local, semi-infinite string operators and requires no enlargement of the Hilbert space. We show that including these operators yields a well-defined SDP even in the thermodynamic limit, and significantly tightens the bounds on thermodynamic ground-state correlations in SSB phases.
This demonstrates the potential for improving bootstrap bounds using operator sets motivated by the excitation spectrum. While we focus mainly on one-dimensional string operators, the ability to incorporate non-local, abstractly defined operators broadens the class of operators that can be included in the bootstrap and may enable the future incorporation of more complicated extended excitations, in higher dimensions or near critical points.

This paper is organized as follows.
We begin with a review of the bootstrap program in \cref{sec:review}.
In \cref{sec:strings}, we motivate the introduction of string operators into the bootstrap program using the example of the 1D TFIM and show that the SDP remains stable in the thermodynamic limit because of the $\mathbb{Z}_2$ symmetry of the model. To establish the generality of the approach beyond the 1D TFIM, we present bounds for several 1D spin models in \cref{sec:results}.
For the 1D TFIM (\cref{sec:TFIM}), we show that the string operators yield significantly tighter bounds in the FM phase.
We then add a next-nearest-neighbor antiferromagnetic interaction to the 1D TFIM to obtain the 1D axial next-nearest-neighbor Ising (ANNNI) model (\cref{sec:annni}).
We show that the bounds remain relatively tight as long as the elementary excitations are captured by the operator set of strings and local operators.
We use the tightness of the bounds across the phase diagram to estimate the phase boundaries where correlations diverge.
We additionally find a region of tight bounds that corresponds precisely to a special line where the Hamiltonian is frustration-free.
In \cref{sec:3clock}, we describe the generalization of the string operators to bootstrap other SSB phases and demonstrate its application to the $\mathbb{Z}_3$ chiral clock (CC$_3$) model.
We conclude in \cref{sec:conclusion}.

\section{Review of the Bootstrap Program}
\label{sec:review}
In this section, we review the construction of a finite bootstrap program by relaxation of constraints obeyed by the ground-state correlations.
In the operator-algebraic sense, a \textit{state} is defined as a positive linear map $\braket{}:\mathcal{B}_{\Lambda}\rightarrow\mathbb{C}$, where $\mathcal{B}_{\Lambda}$ is the algebra of bounded operators acting on the Hilbert space $\mathcal{H}$, and $\mathbb{C}$ is the set of complex numbers.
This definition of state is equivalent to a trace class density matrix which gives the expectations:
\begin{equation}
    \braket{\mathcal{O}} = \text{tr}(\rho \mathcal{O}), \, \forall \mathcal{O}\in \mathcal{B}_{\Lambda}.
\end{equation}

We first define the constraints and their relaxation at finite-size and then demonstrate their generalization to infinite-size.

\subsection{Finite Size}
First, we consider the case of a spin system with a finite number of sites.
The system is specified by a tensor product Hilbert space over a finite set $\Lambda$ of sites, each hosting a $d$-dimensional Hilbert space, and a self-adjoint Hamiltonian $H\in \mathcal{B}_{\mathcal{H}}$.
The ground-state is then defined as a state that satisfies the following constraints \cite{Bratteli1987}:
\begin{enumerate}
    \item Normalization: $\braket{\mathbb{I}} = 1$,
    \item Positivity: $\braket{\mathcal{O}^{\dagger} \mathcal{O}} \geq 0, \, \forall \mathcal{O}\in \mathcal{B}_{\Lambda}$,
    \item Perturbative Positivity: $\braket{\mathcal{O}^{\dagger}[H, \mathcal{O}]} \geq 0, \, \forall \mathcal{O}\in \mathcal{B}_{\Lambda}$.
\end{enumerate}

Here, $\mathbb{I}$ is the identity operator.
Normalization and positivity are the constraints defining a physical state.
Perturbative positivity, on the other hand, excludes states whose energy can be lowered by conjugating with any operator $O\in\mathcal{B}_{\Lambda}$, and thus, defines the ground-state.

A relaxation of the above constraints is obtained by imposing positivity and perturbative positivity only for operators in the span of a finite set $\mathcal{P}\subset \mathcal{B}_{\Lambda}$.
For some self-adjoint $R\in \mathcal{B}_{\Lambda}$, we can define the program:

\begin{equation}
\label{eq:bootstrap}
\begin{split}
    \braket{R}^{\min}_{H, \mathcal{P}} = \text{min}\braket{R},\\
    \text{s.t.}\braket{\mathbb{I}}=1,\\
    \braket{\mathcal{O}^{\dagger}\mathcal{O}}\geq 0, \, \forall \mathcal{O}\in \text{span}(\mathcal{P}),\\
    \braket{\mathcal{O}^{\dagger}[H, \mathcal{O}]} \geq 0, \, \forall \mathcal{O}\in \text{span}(\mathcal{P}).
\end{split}
\end{equation}
$\mathcal{P}$ in turn defines a finite set $\mathcal{Q}\subset \mathcal{B}_{\mathcal{H}}$ of linearly independent, self-adjoint operators such that $p_i^{\dagger} p_j, p_i^{\dagger}[H, p_j]\in \text{span}(Q), \forall p_i, p_j\in \mathcal{P}$ \cite{scheer2024hamiltonian,scheer2025defect}.
In order to get meaningful bounds, we must choose $\mathcal{P}$ such that $R ,\mathbb{I}\in \mathcal{Q}$.
The constraints in \cref{eq:bootstrap} can then be written as a semidefinite program (SDP) over the set of expectation values of operators in $\mathcal{Q}$.

The upper bound $\braket{R}^{\max}_{H, \mathcal{P}}$ can similarly be obtained by running the same optimization program in \cref{eq:bootstrap} for $-R$. In particular, $\braket{R}^{\max}_{H, \mathcal{P}} = -\braket{-R}^{\min}_{H, \mathcal{P}}$. Since the ground-state is necessarily contained in the relaxed set, we obtain rigorous two-sided bounds over all possible ground-state expectations:
\begin{equation}
    \braket{R}^{\min}_{H, \mathcal{P}}\leq \braket{R}\leq \braket{R}^{\max}_{H, \mathcal{P}}.
\end{equation}
Since this method is formulated in terms of the density matrix, we can also impose all the symmetries of the Hamiltonian that leave $R$ invariant to optimize the program \cite{scheer2024hamiltonian,scheer2025defect}.
We use MOSEK \cite{mosek} to solve the symmetrized SDP.

\subsection{Infinite Size}
The thermodynamic limit is taken by considering an increasing sequence $\{\Lambda_n\}$ of finite sets which converge to the infinite set $\Gamma$.
$\mathcal{B}_{\Gamma}$ is then replaced by the quasi-local algebra $\mathcal{A}_{loc}$ which is defined as the norm-completion of $\mathcal{B}_{\Lambda_n}$ for an increasing sequence of finite subsystems $\{\Lambda_n\}$ which converges to the infinite set $\Gamma$ (see Ref. \cite{Bratteli1997} for a more detailed construction).
However, there are subtleties in defining the ground-state in the thermodynamic limit since the Hamiltonian appearing in perturbative positivity is unbounded. 
For appropriately local interactions, the commutator map $\mathcal{O}\rightarrow[H, \mathcal{O}]$ remains bounded and well-defined, and the ground-state can still be defined similarly as in \cref{eq:bootstrap} \cite{Bratteli1997}.

In this work, we shall consider the simpler case of finite-ranged interactions. This assumption ensures that for any local operator $\mathcal{O}$, only a finite number of interactions in the Hamiltonian contribute to $[H, \mathcal{O}]$.
The finite operator set $\mathcal{P}$ can be assumed to be supported on some finite set of sites $\Lambda\subset\Gamma$. 
For short-ranged interactions, only finitely many terms in the Hamiltonian supported in some neighborhood $\Lambda\cup \partial\Lambda$ contribute to the commutator $[H, \mathcal{O}]$ for $\mathcal{O}\in \mathcal{P}$, meaning that the terms in the perturbative positivity constraints stop changing with $n$ when $\Lambda_n\supset\Lambda\cup \partial\Lambda$.

This allows us to define the program as in \cref{eq:bootstrap} to obtain bounds on ground-state correlations that hold on any large enough system $\Lambda'\supset\Lambda\cup\partial\Lambda$, including in the thermodynamic limit. 
For the bounds precisely in the thermodynamic limit, we can also impose translation symmetry on the correlations for a Hamiltonian with translationally symmetric interactions, such that:
\begin{equation}
    \braket{\tau_{\mathbf{r}}(\mathcal{O})} = \braket{\mathcal{O}}, \forall\,  \mathcal{O}, \tau_{\mathbf{r}}(\mathcal{O}) \in \text{span}(\mathcal{Q}), \mathbf{r}\in \Gamma.
\end{equation}
Here, $\tau_{\mathbf{r}}(\mathcal{O})$ is the operator $\mathcal{O}$ translated by $\mathbf{r}$.

\section{Bootstrap with string operators}
\label{sec:strings}

In the previous section, we assumed elements in $\mathcal{P}$ to be supported on a finite set of sites and used this assumption to show that the bootstrap program remained well-defined in the thermodynamic limit. In this section, we show that we can also incorporate genuinely non-local operators while keeping the SDP stable and well-defined in the thermodynamic limit. 
In one-dimensional (1D) systems with symmetry-breaking, incorporating these non-local string operators helps rule out elementary excitations such as domain walls, and are thus crucial in obtaining tight bootstrap bounds.
Importantly, while we focus on 1D models here, the ability to include non-local operators would be crucial in obtaining tight bounds for phases whose elementary excitations extend over long distances (critical phases), or are topological in nature (symmetry-broken or topological phases). 

We use the 1D TFIM to motivate our construction before discussing other 1D models with discrete internal symmetries. 
Consider the Hamiltonian for the 1D TFIM on a lattice $\Gamma$ with open boundaries. 
\begin{equation}
    \label{eq:TFIM}
    H_{\Gamma} = -\sum_{i, i+1 \in \Gamma} Z_i Z_{i+1} -g\sum_{i\in \Gamma} X_i.
\end{equation}
The model has a $\mathbb{Z}_2$ symmetry generated by $\mathbb{X} = \prod_{i\in \Gamma} X_i$.
In the thermodynamic limit, $\Gamma=\mathbb{Z}$, the model shows a spontaneous symmetry breaking phase (SSB) transition at $g = 1$ with a doubly degenerate ferromagnetic/symmetry-broken phase ($g<1$) and a nondegenerate paramagnetic/symmetric phase ($g>1$).

Refs. \cite{Cho2026CoarseGrainedBootstrap, fawzi2024certified} bootstrapped this model in the thermodynamic limit and observed loose bounds for the ground-state energy density and correlation $\braket{Z_0Z_1}$ in the ferromagnetic phase ($g<1$), even after coarse-graining to remarkably large local operator sets in Ref. \cite{Cho2026CoarseGrainedBootstrap}.
In Ref. \cite{scheer2025defect}, we identified that the looseness of these bounds occurs because perturbative positivity constraints with local operators are not strong enough to remove domain wall states.
This can be seen clearly in the $g=0$ limit, where $H_n = -\sum_{i=-n}^{n-1} Z_i Z_{i+1}$ is the classical Ising Hamiltonian. 
Since local operators can only move domain walls without reducing their energy, perturbative positivity is unable to exclude domain wall states, giving loose upper bounds on the energy density and lower bounds on $\braket{Z_0Z_1}$.
Thus, we expect bootstrap bounds to be fundamentally loose in such symmetry broken states due to the presence of domain wall states that haven't been ruled out.

\begin{figure}[htbp]
    \centering
    \includegraphics[width = 0.9\columnwidth]{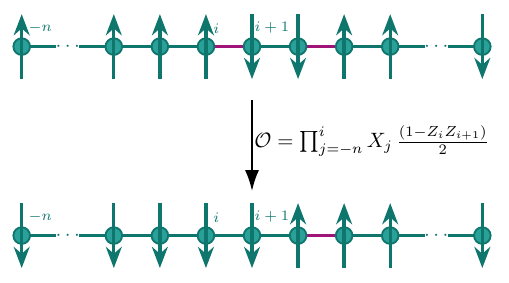}
    \caption{Schematic showing a particular arrangement of spins which has a domain wall across the bond between $i$ and $i+1$. The action of $\mathcal{O}$ on this state removes the domain wall at $i$, and produces a state with lower energy for the Ising Hamiltonian when $g=0$.
    }
    \label{fig:fig1}
\end{figure}

In order to build (non-local) operators that can remove domain walls in the thermodynamic limit, it is constructive to begin with a finite lattice $\Gamma = \Lambda_n = [-n, n]\subset \mathbb{Z}$, and $H_n = H_{\Lambda_n}$, and then take the $n\rightarrow\infty$ limit.
For a finite system $\Lambda_n$, the operator that removes a single domain wall between site $i$ and $i+1$ will be $S^{(n)}_{i} = \prod_{j=-n}^i X_j$ 
, which is a string of spin flips from the left end to site $i$.
As shown in \cref{fig:fig1}, an operator of the form $S^{(n)}_i (1-Z_iZ_{i+1})$ detects a domain wall to the right of site $i$, and removes it by flipping all the spins to the left of site $i+1$.
The $(1-Z_iZ_{i+1})$ factor ensures that the spins are only flipped when there is a domain wall across site $i$.
Since $S^{(n)}_i$ is a truncation of the symmetry operator $\prod_{i\in \Gamma} X_i$, its commutator with the Hamiltonian only receives contributions from interactions between site $j$ and sites to its right:
\begin{equation}
    \label{eq:commrel}
    [H_n, S^{(n)}_j] = 2S^{(n)}_j Z_jZ_{j+1}, \, j<n-1.
\end{equation}
Therefore, at $g=0$, an operator set containing local operators together with  $S^{(n)}_j(1-Z_jZ_{j+1})$ for all $j\in[-n+1, n-1]$ is sufficient to exclude domain-wall states from the bootstrap optimum, giving exact bounds.
Since the excitations close to $g=0$ are also expected to be superpositions of domain walls, we expect the relevant operators to be dressed by some local Paulis and the above operator set to still provide decent bounds for small $g$ in the ferromagnetic phase.

Suppose we take the operator set $\mathcal{P}$ to be some local operators in $\mathcal{B}_{\Lambda_m}$, and their products with a single string operator, say $S_0^{(n)}$, where $n>m$.
For large enough $n$ (concretely $n\geq m+r$ when the Hamiltonian has range-$r$ interactions), the commutator action of the Hamiltonian $H_n$ with the local operators stops changing since all the interaction terms that don't commute with operators in $\mathcal{P}$ have been taken into account with $H_n$. 
Although $S_0^{(n)}$ does depend explicitly on $n$, its commutator action with the Hamiltonian doesn't change since the only contribution to the commutator comes from the $-Z_0Z_1$ interaction as in \cref{eq:commrel}. 
As a consequence, the SDP formulated with an operator set containing local operators in $\mathcal{P}\subset\mathcal{B}_{\Lambda_m}$ along with products of $S_0^{(n)}$ with local operators stops changing for large enough $n$.

This means that the bounds obtained from solving this SDP are valid for ground-state correlations for all systems that contain $\Lambda_{m+r}$, ($r=1$ for the TFIM), including the thermodynamic limit. 
Since we are interested only in the latter, we can also impose translation symmetry on the local correlators to get stronger bounds that are valid only for the infinite-size ground-state. 
Note that the string $S_0^{(n)}$ only rules out domain walls across the bond between $0$ and $1$.
However, imposing translation symmetry applies the corresponding constraints to all translated bonds, effectively removing domain walls throughout the entire chain.

We thus define the string operator $S_0$ in the SDP through its algebra with the local Paulis, using the relations:
\begin{equation}
\label{eq:stringising}
\begin{split}
    S_0^{\dagger}  = S_0, \\
    S_0^2 = \mathbb{I}, \\
    Z_i S_0 = - S_0 Z_i, i\leq 0,\\
    [S_0, Z_i] = 0, i > 0,\\
    [S_0, X_i] = 0, \forall i\in \mathbb{Z}.
\end{split}
\end{equation}

The stability of the SDP with string operators relied on the Hamiltonian being local and having an internal symmetry.
We constructed the string operator by truncating the symmetry operator, which guaranteed stability for a finite-ranged Hamiltonian.
Therefore, we can construct analogous string operators for other symmetries, such as the $\mathbb{Z}_3$ symmetric CC$_3$ model that we shall consider later.

\section{Results}
\label{sec:results}
In this section, we will discuss the results from bootstrapping various 1D models using the relevant string operators.
We find that adding string operators significantly tightens the bounds on ground-state correlations in the SSB phase.

\subsection{1D Transverse Field Ising Model}
\label{sec:TFIM}
\begin{figure}[htbp]
    \centering
    \includegraphics[width = \columnwidth]{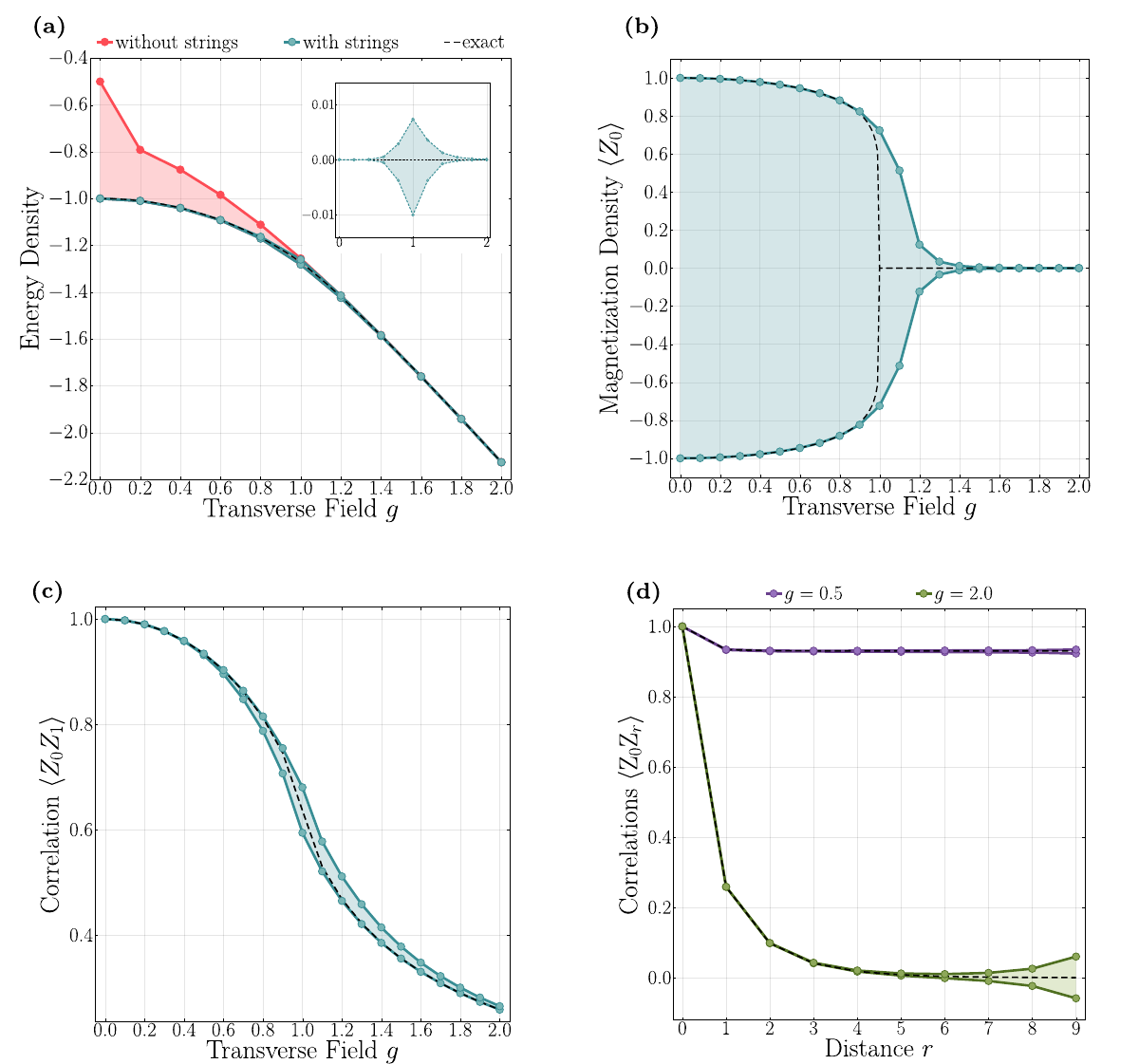}
    \caption{Bootstrap bounds for the 1D TFIM in the thermodynamic limit using string operators. \textbf{(a)} Bounds on the energy density as a function of transverse field, with and without strings. The corresponding operator sets are chosen to be of the same size for a fair comparison. 
    The inset shows the relative difference of the upper and lower bounds with respect to the exact energy density. \textbf{(b)} Bounds on the magnetization density as a function of transverse field. The bounds are symmetric about the origin due to the $\mathbb{Z}_2$ symmetry.
    \textbf{(c)} Bounds on the $\braket{Z_0Z_1}$ correlations as a function of transverse field. 
    \textbf{(d)} Bounds on the long-range correlation $\braket{Z_0Z_r}$ as a function of distance $r$. Tight bounds help us capture the absence and presence of long-range order in the respective phases.
    }
    \label{fig:fig2}
\end{figure}

Here, we present the results from bootstrapping the 1D TFIM in \cref{eq:TFIM} with the string operator, whose algebra is given by \cref{eq:stringising}.
Using a single string $S_0$ and translation symmetry allows us to remove domain walls at every bond in the $g=0$ limit.
However, as shown in \cref{fig:fig2}, we also get tight bounds on the energy density and correlation functions $\braket{Z_0 Z_r}$ for the whole $0<g<1$ ferromagnetic region.
This implies our operator set (products of $S_0$ with local operators) has good overlap with the domain-annihilation operators even for finite transverse fields.
The tight bounds on $\braket{Z_0Z_r}$ for larger $r$ also gives us a diagnostic to detect spontaneous symmetry breaking.

The bounds on magnetization density $\braket{Z_0}$ are also shown in \cref{fig:fig2}.
Since the two ferromagnetic ground-states realize the maximum and minimum values of magnetization density (the degenerate ground-states have the same magnitude of $\braket{Z_0}$ but opposite signs), using the strings to remove domain walls doesn't provide any improvement on these bounds for $g<1$.
Thus, to get tighter magnetization bounds, it is best to use purely local operators in the operator set.

\subsection{Axial Next-Nearest Neighbor Ising Model}
\label{sec:annni}
Next, we add a next-nearest neighbor anti-ferromagnetic coupling $\kappa>0$ to the 1D TFIM, which gives us the 1D axial next-nearest neighbor Ising model (ANNNI) in a transverse field.
\begin{equation}
    H=-\sum_i Z_iZ_{i+1} -g\sum_i X_i +\kappa\sum_i Z_iZ_{i+2}.
\end{equation}

\begin{figure}[htbp]
    \centering
    \includegraphics[width = \columnwidth]{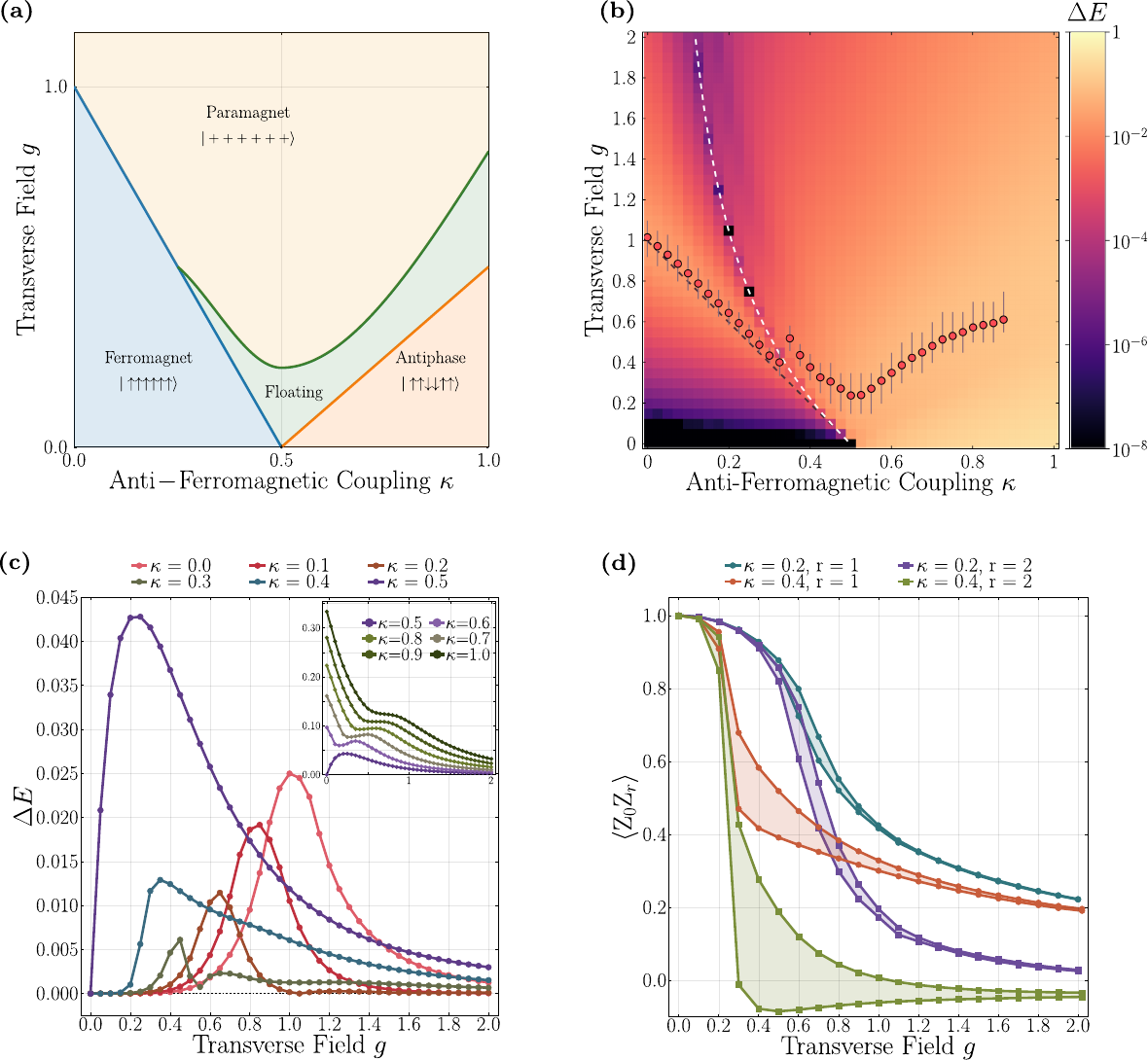}
    \caption{
    \textbf{(a)} Schematic phase diagram of the 1D ANNNI. 
    \textbf{(b)} Uncertainty (difference between upper bound and lower bound) in energy density in the $g-\kappa$ plane. The dots represent the peaks extracted from $\Delta E-g$ cuts at different $\kappa$. The dashed line shows the Peschel-Emery disorder line where the Hamiltonian becomes frustration-free.
    \textbf{(c)} $\Delta E-g$ cuts at different $\kappa\leq0.5$. The inset shows the corresponding cuts for $\kappa\geq 0.5$
    \textbf{(d)} Bounds on the correlation function $\braket{Z_0Z_r}$ as a function of transverse field at two values of $\kappa$.
    The $\kappa = 0.2$ line passes through the FM-PM boundary, whereas the $\kappa = 0.4$ profile also goes through the floating phase. 
    }
    \label{fig:fig3}
\end{figure}

In the classical limit $g=0$, the ANNNI is a canonical frustrated Ising model.
The competition between the nearest-neighbor ferromagnetic coupling and the next-nearest-neighbor antiferromagnetic coupling gives rise to commensurate modulated phases\cite{PhysRevLett.44.1502,selke1988annni}. In the presence of a transverse field, it possesses ferromagnetic, antiphase, paramagnetic, and gapless floating phases \cite{rieger1996one,PhysRevE.75.021105,PhysRevB.76.094410}, as shown in \cref{fig:fig3}.
Unlike the 1D TFIM, the 1D ANNNI is not exactly solvable, except along special lines.

The antiphase is most easily understood in the classical $g=0$ limit.
For $\kappa <0.5$, the ground-state is the doubly degenerate ferromagnet, and the elementary excitations are still the $\mathbb{Z}_2$ Ising domain walls.
For $\kappa>0.5$, the anti-ferromagnetic interaction dominates, favoring anti-aligned next-nearest neighbors.
This is achieved by the period-four pattern $\ket{\uparrow\uparrow\downarrow\downarrow\uparrow\uparrow\downarrow\downarrow...}$, and its translated partners, giving four degenerate ground-states that break translation symmetry and the $\mathbb{Z}_2$ symmetry.
This is known as the (2,2) antiphase order.
In the antiphase, the elementary excitations are discommensurations, which are domain walls between the distinct antiphase patterns, which cannot be removed by the string operator $S_0$.

The transition between the ferromagnet and the antiphase occurs at $\kappa=1/2$ and is first order. 
At this point, known as the \textit{multiphase point}, the ground-state degeneracy becomes exponentially large.
The multiphase point is surrounded by a gapless \textit{floating phase}, as shown in \cref{fig:fig3}.
This is an incommensurate Luttinger liquid phase characterized by algebraically decaying correlations modulated by a continuously varying wave vector.

The operator set consisting of local operators and their products with $S_0$ is tailored to the elementary excitations in the ferromagnetic and paramagnetic regimes: string-dressed operators can remove ferromagnetic domain walls, while local operators capture the paramagnetic excitations. We therefore expect tight bounds deep in both these phases.

We plot the energy-density bound width $\Delta E$ as a function of $(\kappa,g)$ in \cref{fig:fig3}. As expected, the bounds are tight deep in the ferromagnetic and paramagnetic phases. The behavior of $\Delta E$ along fixed-($\kappa$) cuts changes systematically across the phase diagram. For small $\kappa \lesssim 0.25$, the value of $\Delta E(g)$ has a single pronounced maximum at a finite value of $g$, consistent with the location of the FM-PM transition. Near $\kappa \approx 0.3$, the profile begins to develop an additional shoulder or secondary maximum, and for larger $\kappa$ the peak structure becomes broader and more asymmetric. For $\kappa>0.5$, the largest bound width occurs near $g=0$, where the simple ferromagnetic string operator set is no longer adapted to the antiphase ground-state manifold; nevertheless, a finite-$g$ local maximum remains visible up to $\kappa \lesssim 0.8$, after which it smooths into an inflection-like feature.

Beyond the actual bounds on the energy, the gap between the upper and lower bounds provides a useful diagnostic of phase transitions. The reason is that in the vicinity of a continuous phase transition, the correlation length diverges, and a fixed operator set is expected to become less effective in capturing the long-range correlations. 
We thus expect the bound gap, $\Delta E$ to become large around critical points.
This behavior is already visible in our 1D TFIM results, where $\Delta E$ peaks at the critical point as shown in \cref{fig:fig2}.

We apply this analysis to estimate the phase boundaries in the 1D ANNNI model. Here, we expect a similar peak to occur along points on the FM-PM phase boundary, as well as along the floating phase-PM boundary, which is a Kosterlitz-Thouless type transition with a diverging correlation length.
On the other hand, the FM-floating and the antiphase-floating phase transitions fall into the Pokrovsky–Talapov (PT) universality class \cite{PhysRevLett.42.65}, which describes the classical commensurate-incommensurate transitions.
The PT transition point describes the onset of discommensurations in the ground-state, and has a diverging spacing between discommensurations instead of a diverging correlation length.

In \cref{fig:fig3}, we plot the approximate phase boundaries obtained this way by finding the peaks in the bound deviation as a function of $g$ for different values of $\kappa$.
We see that the FM-PM phase boundary for small $\kappa$ follows the expected phase boundary derived using perturbation theory in Ref. \cite{PhysRevE.75.021105}.
For intermediate $\kappa$, we see a jump in the positions of the peaks, which is the expected behavior with the onset of the floating phase.
For $\kappa>0.5$, although we no longer find good bounds in the commensurate (antiphase) region, we can still discern a local peak in $\Delta E$ up to some finite $\kappa>0.5$.
We show the location of this local peak above the antiphase region in \cref{fig:fig3} for the range of $\kappa$ for which a peak is discernible.
These bounds follow the behavior expected from the floating phase-PM phase boundary.

We also see a dark streak of exceptionally tight bounds in the paramagnetic phase in \cref{fig:fig3}. This coincides with the Peschel-Emery (PE) line, along which the quantum ANNNI Hamiltonian has an exactly known product ground state and can be written in a frustration-free form \cite{peschel1981calculation,wouters2021interrelations}. It is therefore natural that a relaxation built from local constraints performs unusually well on this line. This observation is also consistent with previous numerical work identifying the PE line with the disorder line of the quantum ANNNI model, where the character of the paramagnetic correlations changes from monotonic to oscillatory decay \cite{beccaria2006disorder}.

The analysis of the approximate phase boundary and the PE line show the promise of using bootstrap bounds in making quantitative predictions of phase boundaries and frustration in the ground-state. 

\subsection{$\mathbb{Z}_3$ Chiral Clock Model}
\label{sec:3clock}
The above two models had the same $\mathbb{Z}_2$ internal symmetry, and thus, the same $S_0$ described in \cref{eq:stringising}. To show how to generalize the construction beyond $\mathbb{Z}_2$, we now consider the $\mathbb{Z}_p$ clock models \cite{PhysRevB.24.398}.
These models have generated interest due to their connection to $\mathbb{Z}_p$ parafermion chains, which can be used to realize universal quantum computation \cite{fendley2012parafermionic, PhysRevB.90.165106, PhysRevB.92.035154, PhysRevB.93.125105}.

\begin{figure*}[t]
    \centering
    \includegraphics[width = \textwidth]{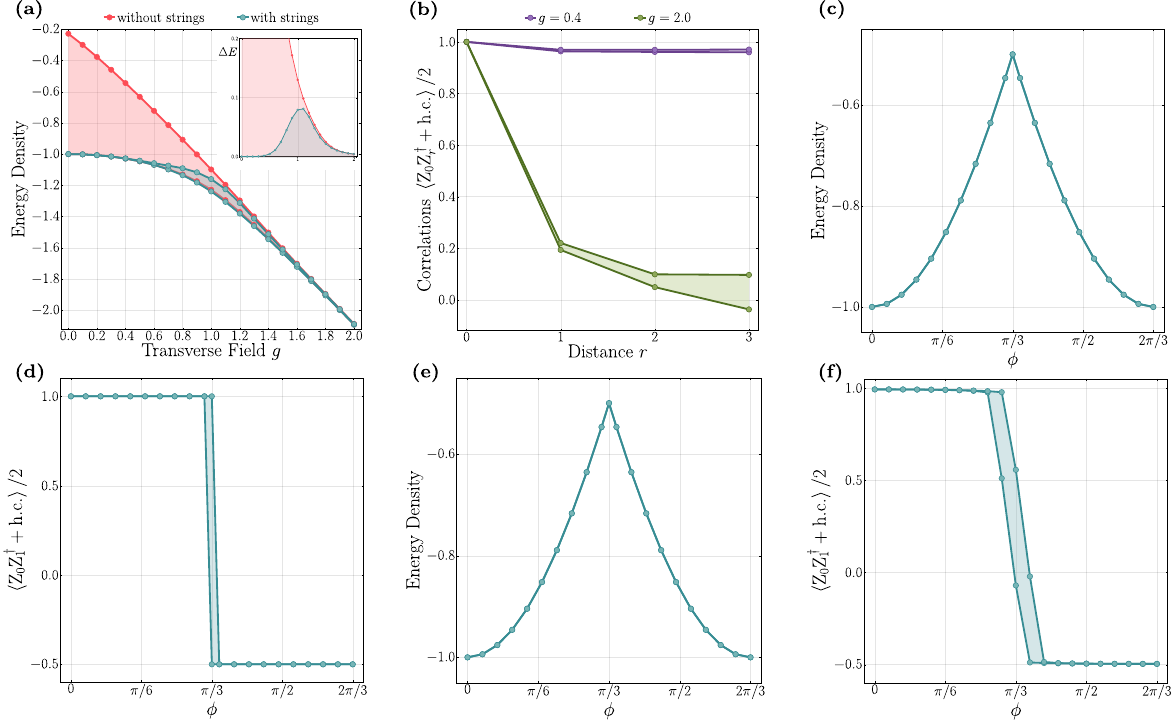}
    \caption{
    Bootstrap bounds for the 1D CC$_3$ in the thermodynamic limit using the string operators. \textbf{(a)} Bounds on the energy density as a function of transverse field, with and without strings. The corresponding operator sets are chosen to be of the same size for a fair comparison. 
    The inset shows the difference between the upper and lower bounds versus the transverse field. \textbf{(b)} Bounds on the correlation functions $\braket{Z_0Z_{r}^{\dagger} + Z_0^{\dagger}Z_r}/2$ as a function of distance for $g=0.4$ and $g=2.0$. 
    \textbf{(c)} and \textbf{(d)} Bounds on the energy density and the nearest neighbor correlation $g(1)$ as a function of $\phi$ for $g=0$. The bounds for energy density in $\textbf{(c)}$ are exact up to floating point precision and hence lie on top of each other.
    }
    \label{fig:fig4}
\end{figure*}

The $\mathbb{Z}_p$ chiral clock model (CC$_p$) is defined on a lattice where each site has an $p$-dimensional subspace.
The generalization of the Pauli operators are the clock and shift operators, whose action on the single-particle basis $\{\ket{m}\}, m=0,1,\dots,p-1$ is defined as:
\begin{enumerate}
    \item Clock $Z$: $Z\ket{m} = \omega^m \ket{m}$ where $\omega= e^{2\pi i/p}$,
    \item Shift $X$: $X\ket{m} = \ket{m+1 \mod p}$.
\end{enumerate}
These operators satisfy the Weyl algebra:
\begin{equation}
\begin{split}
    ZX = \omega XZ,\\
    X^p = Z^p = 1,\\ X^{\dagger} = X^{p-1},\\ Z^{\dagger} = Z^{p-1}.
\end{split}
\end{equation}
For $p=2$, this is isomorphic to the algebra generated by two of the three Pauli matrices, and the model reduces to the 1D TFIM.

We consider here the 1D CC$_3$ model, which corresponds to $p=3$, and is described by the Hamiltonian:
\begin{equation}
\label{eq:3clock}
    H = -\sum_{i\in\Lambda} e^{i\phi}Z_i^{\dagger}Z_{i+1} - g\sum_{i\in\Lambda}e^{i\theta}X_i + h.c.
\end{equation}
The $\mathbb{Z}_3$ symmetry operator for the Hamiltonian is given by $\mathbb{X}_3 = \prod_i X_i$.
In this work, we restrict ourselves to $\theta = 0$, and study the behavior of bounds as a function of $g$ and $\phi$.

Refs. \cite{PhysRevB.24.398, PhysRevB.93.125105} studied the phase diagram of the Hamiltonian in \cref{eq:3clock}, and showed the presence of  gapped phases corresponding to the symmetry-broken (small $g/J$) and the symmetric (large $g/J$) phases as in the 1D TFIM, in addition to more exotic phases.
Turning on $\phi$ in the SSB phase results in a transition to another commensurate SSB phase through a highly degenerate multiphase point \cite{PhysRevB.24.398}.
Similar to the 1D ANNNI, this model also possesses a floating phase surrounding the multiphase point,  parametrized by algebraically decaying correlations and a continuously varying order parameter.

As in the 1D TFIM, directly bootstrapping the above model with local operator constraints gives poor bounds in the ferromagnetic phase due to domain wall excitations.
We thus define the appropriate string operator $S_0$:

\begin{equation}
\label{eq:string3}
\begin{split}
    S_0^{\dagger}  = S_0^{2}, \\
    S_0^3 = \mathbb{I}, \\
    Z_i S_0 = \omega S_0 Z_i, i\leq 0,\\
    [S_0, Z_i] = 0, i > 0,\\
    [S_0, X_i] = 0, \forall i\in \mathbb{Z}.
\end{split}
\end{equation}
A combination of $S_0, S_0^2$, and local operators helps remove domain-wall excitations and provides tight bounds in the ferromagnetic phase, as shown in \cref{fig:fig4}. We use operator sets of similar size while comparing the bounds with and without string operators, and observe that the lower bounds on the energy density are slightly tighter for the purely local operator set. This is because the energy-density lower bounds are largely controlled by local operators. For a fixed operator set size, including product operators involving $S_0$ and $S_0^2$ reduces the number of local operators in the set.

At $g=0$, the CC$_3$ model is classical, and tuning $\phi$ drives first-order transitions between distinct commensurate ground-state orders. These include the ferromagnetic order $\ket{0000\cdots}$ and the two period-three chiral orders $\ket{012012\cdots}$ and $\ket{021021\cdots}$, which are related by reversing the chirality. The string operators $S_0$ and $S_0^2$ are able to remove the corresponding domain walls between locally ordered regions in each of these phases. As a result, the bounds on energy density and correlations shown in \cref{fig:fig4} remain tight across the classical first-order transitions as a function of $\phi$.

This behavior is qualitatively different from what we found in the ANNNI model. There, the string removes the domain walls in the FM phase but not the discommensurations of the antiphase, giving rise to loose bounds beyond $\kappa = 0.5$, even in the classical limit. In the CC$_3$ case at $g=0$, by contrast, the $\mathbb{Z}_3$ strings naturally connect the relevant commensurate orders and exclude the domain-wall excitations across the transition.

At a finite transverse field, the situation changes: the transition between commensurate phases is mediated by an intervening floating phase, as in the 1D ANNNI.
As shown in \cref{fig:fig4}, we obtain loose bounds in a small region around $\phi = \pi/3$ when $g\neq 0$, indicating that our operator set doesn't exclude the excitations in the floating phase.

This construction naturally generalizes to other $\mathbb{Z}_p$ clock models. 
For a given $p$, one can include the non-trivial powers of the truncated symmetry string, $S_0, S_0^2, ... S_0^{p-1}$, along with local operators. 
We expect the strings to be most useful in commensurate symmetry-broken phases, and the local operator constraints to be more important in the paramagnetic regions.

\section{Conclusion}
\label{sec:conclusion}
In conclusion, we have introduced a construction that incorporates non-local string operators into the operator set in order to target domain-wall excitations and improve bootstrap bounds in the SSB phases of 1D systems.
Owing to the underlying symmetry, we show that this construction is stable and can therefore be used to obtain bounds at infinite size.
By adding these string operators to the local operator sets for the 1D TFIM, the 1D ANNNI model in a transverse field, and the 1D CC$_3$ model, we obtain significantly tighter bounds on the energy density and correlations in the ferromagnetic phase. This demonstrates the generality of our approach beyond the 1D TFIM and beyond $\mathbb{Z}_2$ symmetry.

Beyond providing tight bounds within individual phases, we find that the looseness of the bounds serves as a reliable indicator of phase boundaries associated with continuous transitions, where correlations become long-ranged. 
We use this to estimate the phase boundary of the 1D ANNNI model, which agrees well with previous analyses.
In addition, we observe a region of tight bounds in the 1D ANNNI model that corresponds exactly to the frustration-free Peschel-Emery line. We find similar behavior for the CC$_3$ model, where the chiral degree of freedom further allows us to probe other transitions.
We also resolve the first-order transition between distinct commensurate ground-state orders as a function of $\phi$ through the jump in the correlation functions.

It is worth noting that although our construction is motivated by the domain walls of the $g=0$ regime of each model, it provides tight bounds throughout the symmetry-broken regime with $g > 0$. This indicates that our operator sets continue to capture the domain-wall excitations as they become more dressed and delocalized, as long as the character of these excitations does not change qualitatively.

We close by noting that the central feature of our approach is that the relevant excitations can be captured by operators that are non-local but defined algebraically through their relation to local operators, which is what enables a well-defined thermodynamic limit. We anticipate that this strategy can be generalized to address more complicated extended excitations, such as those relevant to symmetry-broken phases in higher dimensions, topologically ordered phases, or gapless and critical phases. We leave such directions to future work.

\begin{acknowledgements}
We thank Da-Chuan Lu, Jonah Herzog-Arbeitman, and Ashvin Vishwanath for fruitful discussions.
This work was supported by the Harvard FAS Dean's Competitive Fund for Promising Scholarship.
The computations involved in this paper were run on the FASRC Cannon Cluster supported by the FAS Division of Science Research Computing Group at Harvard University.

\end{acknowledgements}

\bibliographystyle{apsrev4-2}
\bibliography{references} 

\end{document}